\documentstyle[aps,epsfig,float,amstex]{revtex}

\newcommand{\be}{\begin{equation}}
\newcommand{\ee}{\end{equation}}
\newcommand{\beq}{\begin{eqnarray}}
\newcommand{\eeq}{\end{eqnarray}}

\begin{document}
    
\def\gC{\mbox{\boldmath $C$}}
\def\gZ{\mbox{\boldmath $Z$}}
\def\gR{\mbox{\boldmath $R$}}
\def\gN{\mbox{\boldmath $N$}}
\def\ua{\uparrow}
\def\da{\downarrow}
\def\a{\alpha}
\def\b{\beta}
\def\g{\gamma}
\def\G{\Gamma}
\def\d{\delta}
\def\D{\Delta}
\def\e{\epsilon}
\def\z{\zeta}
\def\h{\eta}
\def\th{\theta}
\def\k{\kappa}
\def\l{\lambda}
\def\L{\Lambda}
\def\m{\mu}
\def\n{\nu}
\def\x{\xi}
\def\X{\Xi}
\def\p{\pi}
\def\P{\Pi}
\def\r{\rho}
\def\s{\sigma}
\def\S{\Sigma}
\def\t{\tau}
\def\f{\phi}
\def\vf{\varphi}
\def\F{\Phi}
\def\c{\chi}
\def\w{\omega}
\def\W{\Omega}
\def\Q{\Psi}
\def\q{\psi}
\def\de{\partial}
\def\inf{\infty}
\def\ra{\rightarrow}
\def\bra{\langle}
\def\ket{\rangle}

\title{Antiferromagnetism  of the $2D$ Hubbard Model at Half Filling:  
Analytic Ground State at Weak Coupling }
\author{Michele Cini and Gianluca Stefanucci}
\address{Istituto Nazionale di Fisica della Materia, Dipartimento di Fisica,\\
Universita' di Roma Tor Vergata, Via della Ricerca Scientifica, 1-00133\\
Roma, Italy}
\maketitle
\begin{abstract}
We introduce a {\em local}   formalism  to deal with the Hubbard  model on a 
$N\times N$ square lattice  (for even $N$) in terms of 
eigenstates of number 
operators, having well defined point symmetry. For $U \rightarrow 0$, 
the low lying shells of the kinetic energy are filled in the ground 
state.  At half filling, using the 
$2N-2$ one-body states of the partially occupied shell ${\cal 
S}_{hf}$,  
 we build   a set of  
$\left(\begin{array}{c} 2N-2 \\ N-1 
\end{array}\right)^{2}$   degenerate unperturbed ground states  with 
$S_{z}=0$ which  are then resolved by the Hubbard interaction 
$\hat{W}=U\sum_{r}\hat{n}_{r\ua}\hat{n}_{r\da}$.  
In  ${\cal 
S}_{hf}$ we study the  many-body 
eigenstates of the kinetic energy  with 
 vanishing eigenvalue of the Hubbard repulsion
($W=0$ states). In the $S_{z}=0$ 
 sector,  this is a $N$ times degenerate multiplet. From the singlet 
 component one  obtains the  ground state of the Hubbard model for 
 $U=0^{+}$, which is unique in agreement with  a 
 theorem by Lieb. The wave function 
 demonstrates an antiferromagnetic order, a lattice step translation 
 being equivalent to a spin flip. We show that the total 
momentum vanishes, while the point symmetry is $s$ or 
$d$ for  even or odd $N/2$, respectively. 
\end{abstract}

\section{Introduction}
\label{intro}

The discovery of high temperature superconductors\cite{bm} enhanced the 
interest in models of  two-dimensional, strongly correlated electron 
systems, such as the $2D$ Hubbard model. The  (repulsive)
 Hubbard hamiltonian   is  highly idealised,
 for istance two particles are allowed to interact only 
 on-site, yet the model  already  displays an interesting phase diagram.  
The fluctuation exchange  (FLEX)\cite{ms} diagrammatic approach, which is based on a conserving
approximation and, independently,  renormalization group techniques\cite{zs}\cite{hm} 
  show that near the antiferromagnetic phase at half filling,
there exists a superconducting phase too, with a momentum dependent 
gap, of $d$ wave symmetry. Indication  of a possible instability of the Fermi
liquid towards 
pairing  near half  filling also comes from cluster diagonalizations\cite{fop}\cite{cibal}\cite{fm}.
Therefore, exact results on the half filled Hubbard model may be relevant 
to antiferromagnetism  and to the mechanism of 
the superconducting instability as well.

In the strong coupling limit the double occupation of the same 
site is energetically suppressed and the model at half filling is 
equivalent to the Heisenberg model with an antiferromagnetic exchange 
interaction\cite{a}. A popular approach takes care of the strong repulsion 
between two opposite spin fermions by a Gutzwiller\cite{Gutzwiller} projection, 
i.e. by  throwing out of the Hilbert space  the  double occupation states.

However, truncating the Hilbert space in this way costs lots of kinetic 
energy, so  at finite $U$ the system must allow double occupation, 
 also in  the ground state. At weak coupling it makes sense to speak 
 about particles in filled shells, which behave much as core electrons 
 in atomic physics, and particles in partially filled, or valence, shells.  
 Remarkably, particles in partially filled 
 shells can \underline{totally} avoid double occupation at \underline {no cost} in energy; they 
 do so, forming {\bf W=0 states,} that are defined as {\bf many-particle eigenstates of the 
 kinetic energy with no double occupation.} Below, using a new 
formalism, we show  how $W=0$ states arise by symmetry. 

An important theorem on the Hubbard model at half filling is due 
to Lieb\cite{l}: the ground state for a bipartite lattice is unique and has spin 
$\frac{1}{2}||{\cal S}_{1}|-|{\cal S}_{2}||$ where $|{\cal S}_{1}|$ ($|{\cal 
S}_{2}|$) is the number of sites 
in the ${\cal S}_{1}$ (${\cal S}_{2}$) sublattice; here and  in the following, 
$|{\cal S}|$ will be the number of elements in the set ${\cal S}$. It 
is worth to observe that the theorem makes no assumptions about the 
symmetry of the lattice. For a $N\times N$ square lattice, with $N$ 
even, the ground state is a singlet and in Ref.\cite{md} it  was shown  
that in the strong coupling limit it has  total 
momentum $K_{tot}=(0,0)$ and $s$ wave ($x^{2}+y^{2}$) or 
$d$ wave ($x^{2}-y^{2}$)  symmetry for even or odd $N/2$ 
respectively.

In this paper we build the exact ground state of the 
Hubbard model at half filling  and weak coupling. In Section 
\ref{hamilteff} we state more precisely the problem we want to solve   
and we define some notations. In Section \ref{localform} we show how for each site 
we can build  a  {\em local} one-body  basis set, which is
well suited to write the antiferromagnetic many-body wave function. 
 Finally in 
Section \ref{symaf} we explore the symmetry properties of the 
ground state, and deduce  the same quantum numbers as  
predicted by Refs.\cite{l}\cite{md} at strong coupling.

The present approach lends itself to obtain exact results  for other 
fillings as well, but this will be shown elsewhere\cite{cps}.

\section{The Hubbard 
Model at Half Filling and Weak Coupling} 
\label{hamilteff}

Let us consider the Hubbard model with hamiltonian 
\begin{equation}
H=H_{0}+\hat{W}=t\sum_{\s}\sum_{\bra r,r'\ket}c^{\dag}_{r\s}c_{r'\s}+
\sum_{r}U\hat{n}_{r\ua}\hat{n}_{r\da},\;\;\;\;U>0,
\label{hamil}
\end{equation}
on a square lattice of $N\times N$ sites with periodic boundary 
conditions and even $N$. Here $\s=\ua,\da$ is the 
spin and $r,\;r'$ the spatial  degrees of freedom of the creation and 
annihilation operators $c^{\dag}$ and $c$ respectively. The sum on 
$\bra r,r'\ket$ is over the pairs of nearest neighbors sites and 
$\hat{n}_{r\s}$  is the number operator on the site 
$r$ of spin $\s$. The point symmetry is 
$C_{4v}$, the Group of a square\cite{appendix}; besides, $H$ is 
invariant under the  commutative Group of Translations ${\mathbf 
T}$ and hence the Space Group\cite{hamer}   ${\mathbf G}={\mathbf 
T} \otimes C_{4v} $; $\otimes$ means the semidirect product. We   
represent sites by $r=(i_{x},i_{y})$ and wave vectors by
$k=(k_{x},k_{y})=\frac{2\p}{N}(i_{x},i_{y})$,
with  $i_{x},i_{y}=0,\ldots,N-1$. In terms of the 
Fourier expanded fermion operators 
$c_{k\s}=\frac{1}{N}\sum_{r}e^{ikr}c_{r\s}$, we have 
$H_{0}=\sum_{k}\e(k)c^{\dag}_{k\s}c_{k\s}$ with $\e(k)=2t
[\cos k_{x}+\cos k_{y}]$. Then the one-body plane wave state $c^{\dag}_{k\s}|0\ket\equiv|k\s\ket$ 
is an  eigenstate of $H_{0}$.

We study the ground state. Let ${\cal S}_{hf}$ denote the set (or shell) of the $k$ wave vectors 
such that $\e(k)=0$. 
At half filling ($N^{2}$ particles) for $U=0$ the ${\cal S}_{hf}$  shell
 is half occupied, while all $|k\ket$ orbitals such that 
$\e(k)<0$ are filled. The $k$ vectors of ${\cal S}_{hf}$ lie on the square having 
vertices $(\pm\pi,0)$ and $(0,\pm\pi)$;
one  readily realizes that the dimension of the 
set ${\cal S}_{hf}$, is $|{\cal S}_{hf}|=2N-2$. Since $N$ is even 
and $H$ commutes with the total spin operators,
\begin{equation}
\hat{S}_{z}=\frac{1}{2}\sum_{r}(\hat{n}_{r\ua}-\hat{n}_{r\da}),\;\;\;\;
\hat{S}^{+}=\sum_{r}c^{\dag}_{r\ua}c_{r\da},\;\;\;\;
\hat{S}^{-}=(\hat{S}^{+})^{\dag},
\label{su2gen}
\end{equation} 
at half filling every ground state of $H_{0}$ is represented in 
the $S_{z}=0$ subspace. Thus, $H_{0}$ has  
$\left(\begin{array}{c} 2N-2 \\ N-1 
\end{array}\right)^{2}$   degenerate unperturbed ground state 
configurations with $S_{z}=0$. We wish to study below how this degeneracy is removed by the 
Coulomb interaction $\hat{W}$ already in first-order perturbation theory. 
Actually most of the degeneracy is removed in first-order, and  with the help of 
Lieb's theorem we shall be  able to single out the true, unique  ground state of $H$.
In Appendix \ref{equivalence} we show that   the structure of the first-order wave functions 
is gained by  diagonalizing $\hat{W}$ in the {\bf  truncated  Hilbert space ${\cal H}$} 
spanned 
by the {\bf states of $N-1$ holes of each spin in ${\cal S}_{hf}$}.
In other terms, one solves a $2N-2$-particle problem in the truncated  
Hilbert space ${\cal H}$ and then, understanding the particles in the 
filled shells, obtains the first-order eigenfunctions of $H$ in the 
full $N^{2}$-particle problem. We underline that the matrix of $H_{0}$ 
in ${\cal H}$ is null,  since 
 by construction ${\cal H}$ is contained in the kernel of $H_{0}$.

 The operator $\sum_{r}\hat{n}_{r\ua}\hat{n}_{r\da}$ has 
  eigenvalues $0,1,2, \ldots$ and so  
 the lowest eigenvalue  of $\hat{W}$ is zero (in other terms,
$\hat{W}$ is positive semi-definite). The  unique
ground state of the Hubbard Hamiltonian for $U=0^{+}$ at half
filling will turn out to be a $W=0$ singlet state of $2N-2$ holes in ${\cal S}_{hf}$ (filled 
shells being understood).  We shall obtain the  $W=0$ states
  $\in{\cal H}$. 
It is clear that, although the $U=0$ case is trivial, at  $U=0^{+}$ we are still 
facing a {\em bona fide}  many-body problem, that we are solving 
exactly\cite{he}.

\section{{\em Local} Formalism for the  ground state wave function}
\label{localform}

 In the present section we first define a basis of {\em local}  orbitals; 
 then, we demonstrate a method for actually 
constructing the basis in the general case. We use the $4 \times 4$ 
case as a simple example and then generalize. Then we show that using 
the local basis, 
the many body wave function of the antiferromagnetic ground state can 
be projected out as the singlet component of a {\bf single} determinant, 
which is amazingly simple for an interacting system.

\subsection{The Basis Set: Definition}
\label{basis}

  Since $\hat{W}$ depends on the occupation number operators $\hat{n}_{r}$, 
it is intuitive that  its 
properties in ${\cal H}$ are best discussed by a suitable one-body 
basis of ${\cal S}_{hf}$ such that at 
least one of these operators is diagonal. In addition, a convenient basis 
should exploit the large ${\mathbf G}$ symmetry of the system. If  ${\cal S}_{hf}$ were a 
complete set ($N^{2}$ states), one would trivially go from plane waves 
to atomic orbitals by a Fourier transformation; instead,
we must  define the local counterparts of plane-wave states
using only the $2N-2$ states that belong to ${\cal S}_{hf}$. 
For each site $r$  we  diagonalize the number operator $\hat{n}_{r}$; 
moreover, since $\hat{n}_{r}$ is compatible with the operations of 
the point symmetry group $C_{4v}$ we also diagonalize the Dirac 
characters of the Group. The set of Dirac characters defines
the irreducible representation ({\it irrep}); thus we write the 
one-body basis 
states $\{|\varphi_{\a}^{(r)}\ket\}$ where  $\a$ comprises the $\hat 
{n}_{r}$ eigenvalue and a $C_{4v}$ irrep label.  However,  different 
sites yield  different sets; the eigenvectors $|\varphi_{\a}^{(0)}\ket$ of $n_{r=0}$ and those 
$|\varphi^{(r)}_{\a}\ket$ of other sites $r$ are connected by  unitary transformations. 
Introducing the primitive translations of the lattice $\hat{e}_{x}=(1,0)$ 
(one step towards the right) 
and $\hat{e}_{y}=(0,1)$ (one step upwards)
the primitive unitary transformations read:
\begin{equation}
|\varphi^{(\hat{e}_{l})}_{\a}\rangle=\sum_{\b=1}^{2N-2}|\varphi_{\b}^{(0)}\rangle
  \bra\varphi_{\b}^{(0)}|\varphi^{(\hat{e}_{l})}_{\a}\rangle
\equiv \sum_{\b=1}^{2N-2}|\varphi_{\b}^{(0)}\rangle 
T_{l_{\b\a}},\;\;\;\;\;l=x,y.
\label{transferT}
\end{equation}
 The translation  matrix $T_{l}$  {\em knows} all the ${\mathbf G}$  symmetry of the 
 system, and will turn out to be very special. Using such a basis set for the half 
filled shell the antiferromagnetic order of the ground state comes 
out in a clear and transparent manner.

 \subsection{Technique for Building the Basis Set }
\label{build}

 To accomplish that, it is not actually necessary to diagonalize any $(2N-2) \times (2N-2)$ 
 matrices. The  number operator 
$\hat{n}_{r}=c^{\dag}_{r}c_{r}$ (for the moment we omit
the spin index) is dealt with most easily by the following

{\bf Theorem:}  {\em Let ${\cal S}$ be an arbitrary set 
of plane-wave eigenstates $\{|k_{i}\ket\}$ of $H_{0}$ and
$(n_{r})_{ij}=\bra k_{i}|\hat{n}_{r}|
k_{j}\ket=\frac{1}{N^{2}}e^{i(k_{i}-k_{j})r}$  the matrix of 
$\hat{n}_{r}$ in ${\cal S}$. 
This  matrix  has 
eigenvalues $\l_{1}=\frac{|{\cal S}|}{N^{2}}$  and
$\l_{2}= \ldots =\l_{|{\cal S}|}=0$
}.

Note that  $|{\cal S}|\leq N^{2}$; if  $|{\cal S}|= N^{2}$
the set is  complete, like the set of all orbitals, and the theorem is 
trivial (a particle sitting on site $r$ is the $n_{r}$ eigenvector 
with eigenvalue 1); otherwise the theorem is an immediate consequence 
of the fact that (see Appendix \ref{detdin})
\begin{equation}
    det|(n_{r})_{ij}-\l\d_{ij}|=(-\l)^{|{\cal S}|-1}
    (\frac{|{\cal S}|}{N^{2}}-\l),\;\;\;\forall r.
    \label{det}
\end{equation}
It is easy to verify that for $r=0$ the 
eigenvector with nonzero eigenvalue is just the totally symmetric 
superposition of all the  $\{|k_{i}\ket\} \in {\cal S}$.

Next, the large set  ${\cal S}_{hf}$ breaks into small pieces if we take full advantage
of the $\mathbf{G}$ symmetry. Any plane-wave state $k$ belongs to a 
one-dimensional irrep of $\mathbf{T}$; moreover, it also belongs to a {\em star} of $k$
vectors connected by operations of $C_{4v}$, and one member of the star 
has $k_{x}\geq k_{y}\geq 0$. We recall that any $k\in {\cal S}_{hf}$  
lies on a square with vertices on the axes at the Brillouin zone 
boundaries.  Choosing an arbitrary $k\in {\cal S}_{hf}$ with
$k_{x}\geq k_{y}\geq 0$, hence $k_{x}+k_{y}=\pi$, the set of vectors $R_{i}k \in {\cal S}_{hf}$,  
where $R_{i}\in C_{4v}$, is a basis for an irrep of $\mathbf{G}$.
The high symmetry vectors $k_{A}=(\pi,0)$ and $k_{B}=(0,\pi)$ 
 are the basis of the only two-dimensional irrep of $\mathbf{G}$, which exists for any $N$.
If $N/2$ is even,  one also finds the high symmetry  wavevectors 
$k=(\pm\pi/2,\pm\pi/2)$ which mix among themselves and yield
a four-dimensional irrep.  In general, when $k$ is not in a special 
symmetry direction,  the vectors $R_{i}k$ are all 
different, so  all the other irreps of $\mathbf{G}$ have dimension 8, 
the number of operations of the point Group $C_{4v}$. 

Below, we shall need the number of these irreps.  Since 8 times the number of 
eight-dimensional irreps + 4 times that of four-dimensional ones + 2 
for the only two-dimensional irrep 
must yield $|{\cal S}_{hf}|=2N-2$, one finds that ${\cal S}_{hf}$ contains 
$N_{e}=\frac{1}{2}(\frac{N}{2}-2)$ irreps of dimension 8 if $N/2$ is even and $N_{o}
=\frac{1}{2}(\frac{N}{2}-1)$ irreps of dimension 8 if $N/2$ is odd.

We note incidentally that $\mathbf{G}$ cannot explain the  degeneracy 
$2N-2$ of ${\cal S}_{hf}$, because the maximum dimension of its irreps 
is 8. Indeed, the {\em accidental} degeneracy of several irreps is due to the presence of  
extra symmetry, i.e. $\mathbf{G}$ is a subgroup of the Optimal Group 
defined in Ref.\cite{ijmp}. 

In this way, ${\cal S}_{hf}$ is seen to be the union of disjoint 
bases of different irreps of the Space Group. This break-up of  ${\cal 
S}_{hf}$ enables us to carry on the analysis and build the basis for 
any $N$.  For illustration, we will first consider the 
case $N=4$ and then generalize.

\subsection{Example: Basis Set and Ground State for the $4\times 4$ Square Lattice}
\label{4case}

As already noted,  $k_{A}=(\p,0)$ and $k_{B}=(0,\p)$ belong to ${\cal S}_{hf}$ 
and are the basis of a two-dimensional irrep  of $\mathbf{G}$. The $2 
\times 2$ matrix  $(n_{r=0})_{ij}=\bra k_{i}|\hat{n}_{r=0}|k_{j}\ket$, with $i,j=A,B$,
has the eigenvector $|\q^{''(0)}_{A_{1}}\ket=\frac{1}{\sqrt{2}}
(|k_{A}\ket+|k_{B}\ket)$ with eigenvalue  $\lambda_{1}=1/8$, in 
agreement with the above theorem. The second eigenvector, with vanishing eigenvalue, is
$|\q^{''(0)}_{B_{1}}\ket=\frac{1}{\sqrt{2}}(|k_{A}\ket-|k_{B}\ket)$. 
As the notation implies, both are simultaneously eigenvectors of the 
Dirac characters and carry symmetry labels; actually the symmetries 
 $A_{1}$ and $B_{1}$ could have been predicted without 
diagonalization because the two-dimensional irrep  of $\mathbf{G}$
breaks into $A_{1}\oplus B_{1}$ in $C_{4v }$.

Translating by $r$, plane wave states pick up a phase factor: $|k\ket\ra e^{ikr} |k\ket$.
Thus, the eigenstates of $n_{r}$ are 
$|\q^{''(r)}_{A_{1}}\ket=\frac{1}{\sqrt{2}}(e^{ik_{A}r}|k_{A}\ket+
e^{ik_{B}r}|k_{B}\ket)$ with $\lambda_{1}=1/8$
and $|\q^{''(r)}_{B_{1}}\ket=\frac{1}{\sqrt{2}}(e^{ik_{A}r}|k_{A}\ket-
e^{ik_{B}r}|k_{B}\ket)$ with $\lambda_{2}=0$. The primitive 
translations (\ref{transferT}) are performed by:
\begin{equation}
|\q^{''(\hat{e}_{l})}_{I}\ket=\sum_{J=A_{1},B_{1}}
|\q^{''(0)}_{J}\ket\bra\q^{''(0)}_{J}|\q^{''(\hat{e}_{l})}_{I}\ket
\equiv \sum_{J=A_{1},B_{1}}|\q^{''(0)}_{J}\ket
(T_{l})_{JI},\;\;\;\;\;l=x,y
\label{transfer}
\end{equation}
with $I=A_{1},B_{1}$. Using eq.(\ref{transfer}) one finds  the 
antidiagonal 
translation matrices $T_{l}$ 
\begin{equation}
T_{x}=\left[\begin{array}{rr}
0 & -1 \\ -1 & 0 \end{array}\right]\;,\;\;\;\;\;\;
T_{y}=\left[\begin{array}{rr}
0 & 1 \\ 1 & 0 \end{array}\right].
\label{transfer2}
\end{equation}
So, the orbital of $A_{1}$ symmetry at $r=0$ has $B_{1}$ symmetry 
around the nearest neighbour sites, and conversely. In 
particular,   $|\q^{''(0)}_{A_{1}}\ket$  has vanishing amplitude on 
a sublattice and $|\q^{''(0)}_{B_{1}}\ket$ on the other. 
The two-body state $|\q^{''(0)}_{A_{1}}\ket_{\s}|\q^{''(0)}_{B_{1}}\ket_{-\s}$
has occupation for spin $\s$ but not for spin $-\s$ on the site $r=0$; 
under a lattice  step  translation it flips the spin and picks up a (-1) phase factor: 
\begin{equation}
|\q^{''(0)}_{A_{1}}\ket_{\s}|\q^{''(0)}_{B_{1}}\ket_{-\s}
\longleftrightarrow 
|\q^{''(0)}_{B_{1}}\ket_{\s}|\q^{''(0)}_{A_{1}}\ket_{-\s}=
-|\q^{''(0)}_{A_{1}}\ket_{-\s}|\q^{''(0)}_{B_{1}}\ket_{\s};
\label{traslx}
\end{equation}
therefore it has double occupation nowhere and is a $W=0$ state (more 
precisely, a $W=0$ pair \cite{cbs1}\cite{cbs2}).

For $N=4$, ${\cal S}_{hf}$ also comprises the basis 
$k_{1}=(\p/2,\p/2),\;k_{2}=(-\p/2,\p/2),\;k_{3}=(\p/2,-\p/2),\; 
k_{4}=(-\p/2,-\p/2)$ of the 4-dimensional irrep of $\mathbf{G}$. This 
irrep  breaks into 
$A_{1}\oplus B_{2}\oplus 
E$ in $C_{4v }$, and such  are the symmetry labels of the eigenvectors 
of $\hat{n}_{r=0}$. We easily obtain them using the  projection 
operators of $C_{4v}$. Letting $I=1,2,3,4$ for the irreps $A_{1},\;B_{2},\;E_{x},\;E_{y}$ 
respectively, we can  write down all the eigenvectors of $\bra k_{i}|\hat{n}_{r=0}|k_{j}\ket$,
with $i,j=1,\ldots,4$, as 
$|\q^{'(0)}_{I}\ket=\sum_{i=1}^{4}O'_{Ii}|k_{i}\ket$, where $O'$ is the 
following $4\times 4$ orthogonal matrix
\begin{equation}
O'=\frac{1}{2}\left[\begin{array}{rrrr}
1 & 1 & 1 & 1 \\
1 & -1 & -1 & 1 \\
1 & -1 & 1 & -1 \\
-1 & -1 & 1 & 1 \end{array}\right].
\end{equation}
The state with non-vanishing eigenvalue is again of $A_{1}$ 
symmetry.  Translating by $r$ we readily get the eigenstates $|\q^{'(r)}_{I}\ket$ of $n_{r}$ and of 
the Dirac characters; in this way, we calculate  the translation matrices $(T_{l})_{JI}=
\bra\q^{'(0)}_{J}|\q^{'(\hat{e}_{l})}_{I}\ket$, which are 
block-antidiagonal: 
\begin{equation}
T_{x}=\left[\begin{array}{rrrr}
0 & 0 & i & 0 \\
0 & 0 & 0 & -i \\
i & 0 & 0 & 0 \\
0 & -i & 0 & 0 
\end{array}\right]\;,\;\;\;\;\;\;
T_{y}=\left[\begin{array}{rrrr}
0 & 0 & 0 & -i \\
0 & 0 & i & 0 \\
0 & i & 0 & 0 \\
-i & 0 & 0 & 0 \\ 
\end{array}\right].
\label{transfer4}
\end{equation}
These $4 \times 4$ translation matrices are again very special; they are 
such that for each lattice step 
the subspace of  $A_{1}$ and $B_{2}$ symmetry is  exchanged with the one  
of $E_{x}$ and $E_{y}$ symmetry, and this means that we are about to 
obtain new $W=0$ states. 
Indeed, $|\q^{'(0)}_{A_{1}}\q^{'(0)}_{B_{2}}\ket_{\s}|\q^{'(0)}_{E_{x}}\q^{'(0)}_{E_{y}}\ket_{-\s}$
is a 4-body eigenstate of $\hat{W}$ with vanishing eigenvalue: 
under a lattice step translation this state does not change its spatial 
distribution but $\s\ra -\s$. We may write that 
\begin{equation}
|\q'_{A_{1}}\q'_{B_{2}}\ket_{\s}|\q'_{E_{x}}\q'_{E_{y}}\ket_{-\s}
\longleftrightarrow
|\q'_{E_{x}}\q'_{E_{y}}\ket_{\s}|\q'_{A_{1}}\q'_{B_{2}}\ket_{-\s}=
|\q'_{A_{1}}\q'_{B_{2}}\ket_{-\s}|\q'_{E_{x}}\q'_{E_{y}}\ket_{\s},
\label{trasldiag}
\end{equation}
and since $\q'_{A_{1}}$ is the only orbital having amplitude at $r=0$ 
it is evident that this state has double occupation nowhere.

At this stage we have distinct local bases for both irreps of
$\mathbf{G}$ having their basis vectors in ${\cal S}_{hf}$.
From eqs.(\ref{transfer2})(\ref{transfer4}) 
we see that under a lattice step translation the subspace spanned by 
$\{|\q^{''(0)}_{A_{1}}\ket$,$|\q^{'(0)}_{A_{1}}\ket, 
|\q^{'(0)}_{B_{2}}\ket\}$ is mapped in the one spanned by 
$\{|\q^{''(0)}_{B_{1}}\ket,|\q^{'(0)}_{E_{x}}\ket,|\q^{'(0)}_{E_{y}}\ket\}$ 
and conversely. Now we are in position to get the exact  ground state 
of the $4\times 4$ square lattice at weak coupling. Consider the 6-body 
determinantal eigenstate of the kinetic term $H_{0}$
\begin{equation}
|\F_{AF}\ket_{\s}=|\q^{''(0)}_{A_{1}}\q^{'(0)}_{A_{1}}\q^{'(0)}_{B_{2}}\rangle_{\s}
|\q^{''(0)}_{B_{1}}\q^{'(0)}_{E_{x}}\q^{'(0)}_{E_{y}}\rangle_{-\s}.
\label{perirrep}
\end{equation} 
In this state there is partial occupation of 
site $r=0$ with spin $\s$, but no double occupation. A
shift by a lattice step produces the transformation
\begin{equation}
  |\F_{AF}\ket_{\s}   \longleftrightarrow - |\F_{AF}\ket_{-\s}
\label{giochino}
\end{equation}
that is, a lattice step is equivalent to a spin flip, a feature that 
we call  {\em antiferromagnetic property}. Since the spin-flipped state is 
also free of double occupation, $|\F_{AF}\ket_{\s}$ is a $W=0$ 
eigenstate, and belongs to the first-order ground-state multiplet. 
 Moreover, the single deteminant with the antiferromagnetic 
property may be analysed in its spin components, which must likewise 
be free of double occupation.  We show below that there is at least one 
$W=0$ state in  ${\cal H}$ for each $S$. By Lieb's theorem, the unique ground state of the Hubbard model 
is the singlet component of (\ref{perirrep}); we shall deal with the 
projection in Sect.\ref{symaf}.

We note that  
$|\q^{''(0)}_{A_{1}}\ket$ and $|\q^{'(0)}_{A_{1}}\ket$ are two 
one-particle states having nonvanishing occupation at $r=0$, and we are getting a new one 
for each irrep of $\mathbf{G}$; for some applications we might prefer  
having only one such state  in ${\cal S}_{hf}$.  This is easily 
accomplished after this 
preparation; shall call $|\varphi\ket$ the new local 
basis.  According to the above theorem,  $\hat{n}_{r}$ has 
eigenvalues $3/8$ and  (5 times) 0.  For $r=0$ the eigenvector of occupation $3/8$ 
is just the totally symmetric superposition of all the $|k\ket$ states 
in ${\cal S}_{hf}$; in terms of the eigenstates introduced above for 
the two-dimensional and four-dimensional irreps of ${\mathbf G}$,
we may write it in the form $|\varphi^{(0)}_{1}\rangle=
\frac{1}{\sqrt{3}}|\q^{''(0)}_{A_{1}}\ket+\sqrt{\frac{2}{3}}
|\q^{'(0)}_{A_{1}}\ket$.  The above theorem also grants that the orthogonal linear combination of
$A_{1}$ eigenstates of $n_{r=0}$, $|\varphi^{(0)}_{2}\rangle=\sqrt{\frac{2}{3}}|\q^{''(0)}_{A_{1}}\ket-
\frac{1}{\sqrt{3}}|\q^{'(0)}_{A_{1}}\ket$
 has 0 eigenvalue.  We are finished with the $\varphi$ basis, because  the remaining local states 
 are just eigenvectors with vanishing eigenvalue, and we may set 
$|\varphi^{(0)}_{3}\ket=|\q^{'(0)}_{B_{2}}\ket$,
$|\varphi^{(0)}_{4}\ket=|\q^{''(0)}_{B_{1}}\ket$, 
$|\varphi^{(0)}_{5}\ket=|\q^{'(0)}_{E_{x}}\ket$ and 
$|\varphi^{(0)}_{6}\ket=|\q^{'(0)}_{E_{y}}\ket$. 
As anticipated,  $|\varphi\ket$ local basis has the advantage that 
$|\varphi^{(0)}_{1}\ket$ is the only element with nonzero occupation 
at $r=0$;  it also preserves the other useful properties that we have 
analyzed above. 
So the $6\times 6$ translation matrix $(T_{l})_{ij}=
\bra\varphi^{(0)}_{i}|\varphi^{(\hat{e}_{l})}_{j}\ket$ is a {\it block 
antidiagonal matrix}:
\begin{equation}
T_{x}=\left[\begin{array}{rrrrrr}
0 & 0 & 0 & -\frac{1}{\sqrt{3}} & i\sqrt{\frac{2}{3}} & 0 \\
0 & 0 & 0 & -\sqrt{\frac{2}{3}} & -\frac{i}{\sqrt{3}} & 0 \\
0 & 0 & 0 & 0 & 0 & -i \\
-\frac{1}{\sqrt{3}} & -\sqrt{\frac{2}{3}} & 0 & 0 & 0 & 0 \\
i\sqrt{\frac{2}{3}} & -\frac{i}{\sqrt{3}} & 0 & 0 & 0 & 0 \\
0 & 0 & -i & 0 & 0 & 0
\end{array}\right]\;,\;\;\;\;\;\;
T_{y}=\left[\begin{array}{rrrrrr}
0 & 0 & 0 & \frac{1}{\sqrt{3}} & 0 & -i\sqrt{\frac{2}{3}} \\
0 & 0 & 0 & \sqrt{\frac{2}{3}} & 0 &  \frac{i}{\sqrt{3}}\\
0 & 0 & 0 & 0 & i & 0 \\
\frac{1}{\sqrt{3}} & \sqrt{\frac{2}{3}} & 0 & 0 & 0 & 0 \\ 
0 & 0 & i & 0 & 0 & 0 \\
-i\sqrt{\frac{2}{3}} &  \frac{i}{\sqrt{3}} & 0 & 0 & 0 & 0
\end{array}\right].
\label{transfer6}
\end{equation}
The $|\varphi\ket$ local basis at any site $r$ splits into the subsets
${\cal S}_{a}=\{|\varphi^{(r)}_{1}\rangle,|\varphi^{(r)}_{2}\rangle,
|\varphi^{(r)}_{3}\rangle\}$, and
${\cal S}_{b}=\{|\varphi^{(r)}_{4}\rangle,|\varphi^{(r)}_{5}\rangle,
|\varphi^{(r)}_{6}\rangle\}$;
a shift by a lattice step sends members of ${\cal S}_{a}$ into linear combinations 
of the members of  ${\cal S}_{b}$, and conversely. For the present $4 
\times 4$ case, we could have obtained the $\varphi$ basis somewhat more 
simply by direct diagonalization on the whole set ${\cal S}_{hf}$, but 
the present approach has the advantage of being viable at large $N$.

 Indeed, $|\varphi^{(0)}_{1}\varphi^{(0)}_{2}\ket$ is 
equivalent to $|\q^{''(0)}_{A_{1}}\q^{'(0)}_{A_{1}}\ket$, because this is just a 
unitary transformation  of the $A_{1}$ wave functions. 
Thus, we may write:
\begin{equation}    
   |\F_{AF}\ket_{\s}=
   |\varphi^{(0)}_{1}\varphi^{(0)}_{2}\varphi^{(0)}_{3}\rangle_{\s}
   |\varphi^{(0)}_{4}\varphi^{(0)}_{5}\varphi^{(0)}_{6}\rangle_{-\s} .
   \label{confi}
\end{equation}

Besides being useful for the sake of illustration because of its 
relative simplicity, the $4 \times 4$ case can be thorougly explored 
on the computer, since the size of  ${\cal H}$ at half filling is 400. 
We have used Mathematica to diagonalize $H+\xi S^{2}$, where a small 
$\xi$ is a numerical device to keep the different spin components of 
the ground state separated. In this way, we {\em observed} the fourfold 
degenerate, $W=0$ ground state which $\xi$ separates into its singlet, 
triplet, quinted and septet components, as expected, with the 
separation growing like $U^{2}$.  The 
antiferromagnetic property of the wave functions was also easily and 
nicely borne out by the numerical results.

\subsection{The $N\times N$ Square Lattice for general Even N}
\label{generalcase}

As discussed in Sect.\ref{build}, we break ${\cal S}_{hf}$ in the 
bases of irreps of  ${\mathbf G}$ it contains. Each basis consists of 
 plane-wave eigenstates $\{|k_{i}\ket\}$ of $H_{0}$ and is converted 
 in a {\em local} one-body basis at site $r$ by diagonalizing 
 $\hat{n}_{r}$ and Dirac's characters . For $N>4$, ${\cal S}_{hf}$ contains $k$ vectors  
that do not possess any special symmetry and we get  eight-dimensional irreps of ${\mathbf G}$  
since  $R_{i}k$ are all different for all $R_{i} \in C_{4v}$.  
 In other terms, any eight-dimensional irrep of ${\mathbf G}$ is the regular representation 
of $C_{4v}$.  Thus, by the Burnside theorem, it breaks  into 
$A_{1}\oplus A_{2}\oplus B_{1}\oplus B_{2}\oplus E\oplus E$, with the 
two-dimensional irrep occurring twice; these are the symmetry labels 
of the local orbitals we are looking for. Now let  ${k\in \cal 
S}_{hf}$. 
The only eigenvector of the matrix $\bra R_{i}k|\hat{n}_{r=0}|R_{j}k\ket$
corresponding to the non-vanishing eigenvalue $\lambda$ belongs to 
$A_{1}$. Let 
$R_{i},\;i=1,\ldots,8$ denote respectively the identity $\mathbf{1}$, the 
counterclockwise and clockwise 90 degrees rotation
 $C_{4}^{(+)},\;C_{4}^{(-)}$, 
the  180 degrees rotation $C_{2}$, the reflection with respect 
to the $y=0$ and $x=0$ axis $\s_{x},\;\s_{y}$ 
and the reflection with respect to the 
$x=y$ and $x=-y$ diagonals $\s'_{+},\;\s'_{-}$.  
We can write down the eigenvectors of 
the above $n_{r=0}$ matrix as 
$|\q^{(0)}_{I}\ket=\sum_{i=1}^{8}O_{Ii}|R_{i}k\ket$, where $k_{x}\geq 
k_{y}\geq 0$ and  $O$ is the 8$\times$8 orthogonal matrix 
\begin{equation}
O=\frac{1}{\sqrt{8}}
\left[\begin{array}{rrrrrrrr}
                              1 & 1 & 1 & 1 & 1 & 1 & 1 & 1 \\
			      1 & -1 & -1 & 1 & -1 & -1 & 1 & 1 \\
			      1 & 1 & -1 & -1 & 1 & -1 & -1 & 1 \\
			      1 & -1 & 1 & -1 & -1 & 1 & -1 & 1 \\
                              1 & 1 & 1 & 1 & -1 & -1 & -1 & -1 \\
			      1 & -1 & -1 & 1 & 1 & 1 & -1 & -1 \\
			      1 & -1 & 1 & -1 & 1 & -1 & 1 & -1 \\
			      -1 & -1 & 1 & 1 & 1 & -1 & -1 & 1 
\end{array}\right].
\label{ort}
\end{equation}
Here, denoting by $E^{\prime}$ the second occourrence of the irrep $E$,
$I=1,\ldots,8$  is the
$A_{1}$, $B_{2}$, $E_{x}$, $E_{y}$, $A_{2}$, 
$B_{1}$, $E^{\prime}_{x}$, $E^{\prime}_{y}$ 
irrep respectively. A translation by $r$ yields the eigenstates $|\q^{(r)}_{I}\ket$ 
of $\bra R_{i}k|\hat{n}_{r}|R_{j}k\ket$ and of the Dirac characters of the 
point symmetry group. After very long but elementary algebra 
one finds that the translation matrices $(T_{l})_{JI}=
\bra\q^{(0)}_{J}|\q^{(\hat{e}_{l})}_{I}\ket$ are
\begin{equation}
T_{x}=\left[\begin{array}{llllllll}
0 & 0 & 0 & 0 & 0 & \cos k_{x} & i\sin k_{x} & 0 \\
0 & 0 & 0 & 0 & \cos k_{x} & 0 & 0 & -i\sin k_{x} \\
0 & 0 & 0 & 0 & 0 & i\sin k_{x} & \cos k_{x} & 0 \\
0 & 0 & 0 & 0 & i\sin k_{x} & 0 & 0 & -\cos k_{x} \\
0 & \cos k_{x} & 0 & i\sin k_{x} & 0 & 0 & 0 & 0 \\
\cos k_{x} & 0 & i\sin k_{x} & 0 & 0 & 0 & 0 & 0 \\
i\sin k_{x} & 0 & \cos k_{x} &  0 & 0 & 0 & 0 & 0 \\
0 & -i\sin k_{x} & 0 & -\cos k_{x} & 0 & 0 & 0 & 0 \\
\end{array}\right]
\end{equation}
and
\begin{equation}
T_{y}=\left[\begin{array}{llllllll}
0 & 0 & 0 & 0 & 0 & -\cos k_{x} & 0 & -i\sin k_{x} \\
0 & 0 & 0 & 0 & -\cos k_{x} & 0 & i\sin k_{x} & 0 \\
0 & 0 & 0 & 0 & i\sin k_{x} & 0 & -\cos k_{x} & 0 \\
0 & 0 & 0 & 0 & 0 & i\sin k_{x} & 0 & \cos k_{x} \\
0 & -\cos k_{x} & i\sin k_{x} & 0 & 0 & 0 & 0 & 0 \\
-\cos k_{x} & 0 & 0 & i\sin k_{x} & 0 & 0 & 0 & 0 \\
0 & i\sin k_{x} & -\cos k_{x} & 0 & 0 & 0 & 0 & 0 \\
-i\sin k_{x} & 0 & 0 & \cos k_{x} & 0 & 0 & 0 & 0
\end{array}\right]
\end{equation}
where we have taken into account that at half filling 
$k_{y}=\p-k_{x}$. As for the two and four-dimensional irreps the 
translation matrices are in an {\it antidiagonal block form}; in 
particular they are such that the set containing the irreps 
$A_{1},\;B_{2},\;E$ is mapped in the set containing the
irreps $A_{2},\;B_{1},\;E'$ and vice versa. This means that if we put 4 
particles with spin $\s$ in the former 4 irreps and 4 particles with spin 
$-\s$ in the latter 4 ones we obtain an 8-body state for which the 
translation by a lattice step is exactly equivalent to a spin-flip  ({\em antiferromagnetic property}):
\begin{equation}
|\q^{(0)}_{A_{1}}\q^{(0)}_{B_{2}}\q^{(0)}_{E_{x}}\q^{(0)}_{E_{y}}\ket_{\s}
|\q^{(0)}_{A_{2}}\q^{(0)}_{B_{1}}\q^{(0)}_{E'_{x}}\q^{(0)}_{E'_{y}}\ket_{-\s}
\longleftrightarrow
|\q^{(0)}_{A_{2}}\q^{(0)}_{B_{1}}\q^{(0)}_{E'_{x}}\q^{(0)}_{E'_{y}}\ket_{\s}
|\q^{(0)}_{A_{1}}\q^{(0)}_{B_{2}}\q^{(0)}_{E_{x}}\q^{(0)}_{E_{y}}\ket_{-\s}.
\label{traslstella}
\end{equation}

Now recall that the occupation number 
vanishes for all the {\it local} states except the one of symmetry 
$A_{1}$.  Since in each site the $A_{1}$ state of spin $\s$ does not have the 
partner of the same symmetry with spin $-\s$, the 8-body state of 
eq.(\ref{traslstella}) cannot 
have double occupancy on any site and therefore it is a $W=0$ state.

From now on we shall be engaged with  the explicit construction of 
many-body $W=0$ states at half filling and it will be understood that we are using the local basis  
of the site   $r=0$,   so that $|\q^{(0)}\ket\equiv|\q\ket$.  We have 
observed above how the {\em antiferromagnetic property} of a 
many-particle determinant  ensures that it is a $W=0$ state. In this 
way, we easily obtain a determinantal ground state of $\hat{W}$ in ${\cal H}$, i.e., at half 
filling, by creating holes in all the local orbitals of all the  irreps, half with spin up and half with 
spin down.  Let  $|\q_{I}^{[m]}\ket$  be  the one-body eigenstate  of $n_{r=0}$ 
belonging to the irrep $I$ of $C_{4v}$, in the space spanned by  the basis functions of the
$m$-th eight-dimensional irrep of ${\mathbf G}$. For even $N/2$,the 
four-dimensional representation of ${\mathbf G}$ exists and the $W=0$ state 
wave function for the half filled case  
\begin{equation}
|\F_{AF}\ket_{\s}\equiv|(\prod_{m=1}^{N_{e}}
\q^{[m]}_{A_{1}}\q^{[m]}_{B_{2}}\q^{[m]}_{E_{x}}\q^{[m]}_{E_{y}})
\q'_{A_{1}}\q'_{B_{2}}\q''_{A_{1}}
\ket_{\s}
|(\prod_{m=1}^{N_{e}}\q^{[m]}_{A_{2}}\q^{[m]}_{B_{1}}\q^{[m]}_{E'_{x}}
\q^{[m]}_{E'_{y}})\q'_{E_{x}}\q'_{E_{y}}\q''_{B_{1}}\ket_{-\s},
\label{detaf}
\end{equation}
with $\s=\ua,\da$ belongs to the first-order ground state multiplet  
(filled shells are understood, of course). 
For odd $N/2$, on the other hand,
\begin{equation}
|\F_{AF}\ket_{\s}\equiv|(\prod_{m=1}^{N_{o}}
\q^{[m]}_{A_{1}}\q^{[m]}_{B_{2}}\q^{[m]}_{E_{x}}\q^{[m]}_{E_{y}})
\q''_{A_{1}}
\ket_{\s}
|(\prod_{m=1}^{N_{o}}\q^{[m]}_{A_{2}}\q^{[m]}_{B_{1}}\q^{[m]}_{E'_{x}}
\q^{[m]}_{E'_{y}})\q''_{B_{1}}\ket_{-\s}.
\label{detafodd}
\end{equation}

We can see from eqs.(\ref{traslx})(\ref{trasldiag})(\ref{traslstella}) that 
$|\F_{AF}\ket_{\s}$ flips the spin and picks up a phase factor (see 
below) for each lattice step translation.  Therefore, it manifestly shows an 
antiferromagnetic order ({\em antiferromagnetic property}). These 
results generalize Equation (\ref{perirrep}). In both cases, the $-\s$ 
orbitals belong to $n_{r=0}=0$, and the antiferromagnetic property 
grants that the states are $W=0$. Equivalently, we could 
have obtained  a generalised version of (\ref{confi}) by building a 
basis of symmetry adapted eigenvectors of $\hat{n}_{r}$  on the 
whole set ${\cal S}_{hf}$, which can be done without handling large 
matrices.

A few further remarks about $|\F_{AF}\ket_{\s}$ are in order.  
1) Introducing the projection operator 
$P_{S}$  on the spin $S$ subspace, one finds that  
$P_{S}|\F_{AF}\ket_{\s}\equiv|\F^{S}_{AF}\ket_{\s}\neq 0,\;\forall S=0,\ldots,N-1$. 
Then, $_{\s}\bra \F_{AF}|\hat{W}|\F_{AF}\ket_{\s}=\sum_{S=1}^{N-1}\, 
_{\s}\bra\F^{S}_{AF}|\hat{W}|\F^{S}_{AF}
\ket_{\s}=0$, and this implies that there is at least one  $W=0$ state 
of $\hat{W}$ in  ${\cal H}$ for each $S$.
The  ground state of $H$ at weak coupling 
is the singlet $|\F^{0}_{AF}\ket_{\s}$. 2) 
The {\em existence} of this singlet $W=0$ ground state 
is also a direct consequence of the Lieb theorem\cite{l}. Indeed 
the maximum spin state $|\F^{N-1}_{AF}\ket_{\s}$ is trivially in the kernel
of $\hat{W}$; since the 
ground state must be a singlet it should be an eigenvector of $W$ 
with vanishing eigenvalue.  3) The above results and 
Lieb's theorem imply that second and
higher order effects split the ground state multiplet of $H$
and the singlet is lowest.  4) The 
  Lieb  theorem makes no assumptions concerning the lattice 
structure; adding the ingredient of the $\mathbf{G}$ symmetry we are able 
to explicitly display the wave function at weak coupling.

In the next section we study the symmetries properties of the singlet 
component of $|\F_{AF}\ket_{\s}$.

\section{Spin projection and Symmetries of $|\F_{AF}\ket$}
\label{symaf}

The  $W=0$ state $|\F_{AF}\ket_{\s}$ is a $2(N-1)$-body
determinantal state with $S_{z}=0$ but is not an 
eigenstate of the total spin operator $\hat{S}^{2}$. 
The various spin components are degenerate in first-order 
perturbation theory, but when higher-order effects are allowed the 
singlet component is lowest; if we wish to study the ground state of 
the Hubbard model we must project on the singlet. The spin projection operators $P_{S}$ are well 
known and are reviewed in Appendix \ref{spin} for the sake of clarity. 
In order to find out the good quantum numbers of the ground state, the antisymmetric form 
\begin{equation}
  |\F_{AF}\ket\equiv \frac{|\F_{AF}\ket_{\s}-
|\F_{AF}\ket_{-\s}}{\sqrt{2}} 
    \label{report}
\end{equation}
is more convenient to work with than the single determinant   
$|\F_{AF}\ket_{\s}$; using  the explicit form 
of $P_{S=0}$ one finds that the projection is the same.
\subsection{symmetry under translations}

Eqs.(\ref{traslx})(\ref{trasldiag})(\ref{traslstella}) tell us that 
under a lattice step translation each of the determinantal states $|\F_{AF}\ket_{\s}$
of eqs.(\ref{detaf},\ref{detafodd}) undergoes a spin flip, which does 
not change the irreps of $C_{4v}$ but modifies the order in which 
they appear in the many-body state. Since the fermion operators 
anticommute, the translated determinant is 
$(-1)^{N-1}|\F_{AF}\ket_{-\s}=-|\F_{AF}\ket_{-\s}$; 
but in view of Equation  (\ref{report}), 
$|\F_{AF}\ket\ra |\F_{AF}\ket$ under a lattice step translation. Thus  
$|\F_{AF}\ket$ is an eigenstate of the total momentum with 
eigenvalue $K_{tot}=(0,0)$.  Since 
the spin projection cannot change this quantum number, it holds 
for $|\F^{S=0}_{AF}\ket$ too.

\subsection{reflections and rotations}
Now we study how $|\F_{AF}\ket$ transforms under 
reflections and rotations with respect to the center of an 
arbitrary plaquette of the square lattice. We are not compelled to 
refer the operations to  the center of a plaquette, rather than to a site, 
to characterize the  symmetry 
properties of $|\F_{AF}\ket$; indeed the system is $C_{4v}$ invariant in both 
cases. The only reason is to make contact with Ref.\cite{md}. 

Since we represent sites by $r=(i_{x},i_{y})$ with 
$i_{x},i_{y}=0,\ldots,N-1$, we may choose the center at $r_{plaq}=(1/2,1/2)$.
Let $R^{plaq}_{i}$ denote the $C_{4v}$ operations with 
respect to $r_{plaq}$ and $R_{i}$ the ones with respect to the origin 
$(0,0)$. Then for every vector $r$ of our lattice we have from 
elementary geometry
\begin{equation}
R_{i}^{plaq}r=R_{i}(r-r_{plaq})+r_{plaq}.
\end{equation}
This implies the transformation law for plane-wave states   $|k\ket$: 
\begin{equation}
|k\ket=\frac{1}{N}\sum_{r}e^{-ikr}|r\ket\ra 
|R_{i}^{plaq}k\ket
\equiv\frac{1}{N}\sum_{r}e^{-ikr}|R^{plaq}_{i}r\ket=
 e^{-ik(r_{plaq}-R^{-1}_{i}r_{plaq})}|R_{i}k\ket,
\label{rotrifsim}
\end{equation}
where the last equality can be obtained with a change of variables.
By means of eq.(\ref{rotrifsim}) it is possible to know how each irrep 
of the space group $\mathbf{G}$ transforms.

\underline{Two-dimensional irrep:}
Let us first 
consider the two-dimensional irrep whose basis vectors are 
$|\q''_{A_{1}}\ket=\frac{1}{\sqrt{2}}
(|k_{A}\ket+|k_{B}\ket)$ 
and $|\q''_{B_{1}}\ket=\frac{1}{\sqrt{2}}(|k_{A}\ket-|k_{B}\ket)$. 
Under $R^{plaq}_{i}$ we have 
\begin{eqnarray}
|\q''_{A_{1}}\ket\ra|R^{plaq}_{i}\q''_{A_{1}}\ket=
\frac{1}{\sqrt{2}}(|R^{plaq}_{i}k_{A}\ket+|R^{plaq}_{i}k_{B}\ket),\nonumber \\
|\q''_{B_{1}}\ket\ra|R^{plaq}_{i}\q''_{B_{1}}\ket=
\frac{1}{\sqrt{2}}(|R^{plaq}_{i}k_{A}\ket-|R^{plaq}_{i}k_{B}\ket).
\end{eqnarray}
The transformed {\it local} states $|R^{plaq}_{i}\q''_{I}\ket$, with 
$I=A_{1},B_{1}$, can be expressed in terms of the original ones 
$|\q''_{I}\ket$: 
\begin{equation}
|R^{plaq}_{i}\q''_{I}\ket=\sum_{J=A_{1},B_{1}}|\q''_{J}\ket
\bra\q''_{J}|R^{plaq}_{i}\q''_{I}\ket
\label{symm2}
\end{equation}
Using eqs.(\ref{rotrifsim}) and computing the overlaps  
$\bra\q''_{J}|R^{plaq}_{i}\q''_{I}\ket$ of eq.(\ref{symm2}) 
we have studied how the $W=0$ pair state in 
eq.(\ref{traslx}) transforms under a 90 degrees rotation 
$C_{4}^{(+)}$ and a reflection with respect to the $y=0$ and $x=y$ 
axis $\s_{x}$ and $\s'_{+}$ respectively. 
After some algebra it is possible to show that
\begin{equation}
    \begin{array}{ll}
|\q''_{A_{1}}\ket_{\s}|\q''_{B_{1}}\ket_{-\s}\longleftrightarrow
-|\q''_{B_{1}}\ket_{\s}|\q''_{A_{1}}\ket_{-\s}&\;\;\;\;C_{4}^{(+)} \\
 & \\
|\q''_{A_{1}}\ket_{\s}|\q''_{B_{1}}\ket_{-\s}\longleftrightarrow
|\q''_{B_{1}}\ket_{\s}|\q''_{A_{1}}\ket_{-\s}&\;\;\;\;\s_{x} \\
 & \\ 
|\q''_{A_{1}}\ket_{\s}|\q''_{B_{1}}\ket_{-\s}\longleftrightarrow
-|\q''_{A_{1}}\ket_{\s}|\q''_{B_{1}}\ket_{-\s}&\;\;\;\;\s'_{+}.
\end{array}
\label{rr2}
\end{equation}
where $\longleftrightarrow$ means that the left hand side transforms 
in the right hand side and conversely under the operation specified on 
the right.

For a $2\times 2$ square lattice the two-dimensional irrep is the only one in ${\cal S}_{hf}$ 
and $|\F_{AF}\ket$ is explicitly given by 
\begin{equation}
  |\F_{AF}\ket=|\q''_{A_{1}}\ket_{\s}|\q''_{B_{1}}\ket_{-\s}+
  |\q''_{B_{1}}\ket_{\s}|\q''_{A_{1}}\ket_{-\s}.
\end{equation}
By means of eqs.(\ref{rr2}) is not hard to see that $|\F_{AF}\ket$ 
transforms as a $d$ wave of $x^{2}-y^{2}$ symmetry. This symmetry property 
cannot change after the spin projection of $|\F_{AF}\ket$ on the $S=0$ 
subspace and we conclude that $|\F^{S=0}_{AF}\ket$ transforms as a $d$ 
wave too. Finally we observe the the $2\times 2$ case is special 
because $|\F_{AF}\ket$ is already a singlet and then must coincide 
with $|\F^{S=0}_{AF}\ket$. This will be no more true if $N\geq 4$. 

\underline{Four-dimensional irrep:} 
Similarly we can study the behaviour of the four-body $W=0$ state of 
eq.(\ref{trasldiag}) under the three operations in eq.(\ref{rr2}). 
One finds:
\begin{equation}
    \begin{array}{ll}
|\q'_{A_{1}}\q'_{B_{2}}\ket_{\s}|\q'_{E_{x}}\q'_{E_{y}}\ket_{-\s}
\longleftrightarrow-|\q'_{E_{x}}\q'_{E_{y}}\ket_{\s}
|\q'_{A_{1}}\q'_{B_{2}}\ket_{-\s}&\;\;\;\;C_{4}^{(+)} \\ & \\
|\q'_{A_{1}}\q'_{B_{2}}\ket_{\s}|\q'_{E_{x}}\q'_{E_{y}}\ket_{-\s}
\longleftrightarrow |\q'_{E_{x}}\q'_{E_{y}}\ket_{\s}
|\q'_{A_{1}}\q'_{B_{2}}\ket_{-\s}&\;\;\;\;\s_{x}\\ & \\
|\q'_{A_{1}}\q'_{B_{2}}\ket_{\s}|\q'_{E_{x}}\q'_{E_{y}}\ket_{-\s}
\longleftrightarrow -
|\q'_{A_{1}}\q'_{B_{2}}\ket_{\s}|\q'_{E_{x}}\q'_{E_{y}}\ket_{-\s}
&\;\;\;\;\s'_{+}.\end{array}
\label{rr4}
\end{equation}

A remarkable feature follows from the above transformation 
properties. In the $N=4$ case ${\cal S}_{hf}$ contains the 
irrep of dimension 2 and the one of dimension 4 so that 
$|\F_{AF}\ket$ is
\begin{equation}
  |\F_{AF}\ket=|\q'_{A_{1}}\q'_{B_{2}}\q''_{A_{1}}\ket_{\s}
  |\q'_{E_{x}}\q'_{E_{y}}\q''_{B_{1}}\ket_{-\s} +
  |\q'_{E_{x}}\q'_{E_{y}}\q''_{B_{1}}\ket_{\s}
  |\q'_{A_{1}}\q'_{B_{2}}\q''_{A_{1}}\ket_{-\s}.
\end{equation}
Using eqs.(\ref{rr2})(\ref{rr4}) it can be shown that 
$|\F_{AF}\ket$ transforms as an $s$ wave of $x^{2}+y^{2}$ symmetry 
for $N=4$. Therefore 
the symmetry of $|\F_{AF}\ket$ depends on $N$ and the next step 
will be to determine how this happens. To this end we need the transformation properties of an arbitrary 
8-dimensional irrep of $\mathbf{G}$ contained in ${\cal S}_{hf}$. 

\underline{Eight-dimensional irrep:} 

After very long but simple algebra we have found that the 
eight-body state of eq.(\ref{traslstella}) transforms as 
\begin{equation}
    \begin{array}{ll}
|\q_{A_{1}}\q_{B_{2}}\q_{E_{x}}\q_{E_{y}}\ket_{\s}
|\q_{A_{2}}\q_{B_{1}}\q_{E'_{x}}\q_{E'_{y}}\ket_{-\s}\longleftrightarrow
|\q_{A_{2}}\q_{B_{1}}\q_{E'_{x}}\q_{E'_{y}}\ket_{\s}
|\q_{A_{1}}\q_{B_{2}}\q_{E_{x}}\q_{E_{y}}\ket_{-\s}&C_{4}^{(+)}
\\ & \\
|\q_{A_{1}}\q_{B_{2}}\q_{E_{x}}\q_{E_{y}}\ket_{\s}
|\q_{A_{2}}\q_{B_{1}}\q_{E'_{x}}\q_{E'_{y}}\ket_{-\s}\longleftrightarrow
|\q_{A_{2}}\q_{B_{1}}\q_{E'_{x}}\q_{E'_{y}}\ket_{\s}
|\q_{A_{1}}\q_{B_{2}}\q_{E_{x}}\q_{E_{y}}\ket_{-\s} &\s_{x}
\\ & \\
|\q_{A_{1}}\q_{B_{2}}\q_{E_{x}}\q_{E_{y}}\ket_{\s}
|\q_{A_{2}}\q_{B_{1}}\q_{E'_{x}}\q_{E'_{y}}\ket_{-\s}\longleftrightarrow
|\q_{A_{1}}\q_{B_{2}}\q_{E_{x}}\q_{E_{y}}\ket_{\s}
|\q_{A_{2}}\q_{B_{1}}\q_{E'_{x}}\q_{E'_{y}}\ket_{-\s} &\s'_{+}.
\end{array}
\label{rr8}
\end{equation}
 For $k_{x}\ra \p/2$ the 
8-dimensional irrep is equivalent to two 4-dimensional ones and the 
eqs.(\ref{rr8}) are the ``square'' of the eqs.(\ref{rr4}). Analogously 
for $k_{x}\ra \p$ we obtain four times the 2-dimensional irrep and 
eqs.(\ref{rr8}) are the ``fourth power'' of eqs.(\ref{rr2}). 

From (\ref{rr8}) we deduce that whatever is the number of 8-dimensional 
irreps, the symmetry of $|\F_{AF}\ket$ depends only by the presence 
or absence of the 4-dimensional one. More exactly if $N/2$ is even 
$|\F_{AF}\ket$ belongs to the one-dimensional irrep $A_{1}$  
and if $N/2$ is odd to the one-dimensional irrep $B_{1}$. 

As noted at the beginning of this section, the  
spin projection on the singlet subspace does not alter the above 
quantum numbers. Therefore we conclude that the $W=0$ singlet state 
$|\F^{S=0}_{AF}\ket$ has total momentum $K_{tot}=(0,0)$ and transforms 
as a $d$ wave of $x^{2}-y^{2}$ symmetry if $N/2$ is odd and as an $s$ 
wave of $x^{2}+y^{2}$ symmetry if $N/2$ is even. The same quantum numbers 
were obtained in Ref.\cite{md} for the ground state of the 
Hubbard model in the opposite, strong coupling regime. 
This coincidence is  a further 
consequence of Lieb's theorem. Since the ground state at half filling must be 
unique, no level crossing is allowed for finite $U$, and the symmetry 
of the ground state is the same at weak and strong coupling.
\section{Acknowledgements}
This work was supported by the Istituto Nazionale di Fisica della 
Materia.

\appendix
\section{Contributions to the $\hat{W}$ matrix from Filled Shells }
\label{equivalence}

The $ N^{2}-$body determinantal wave functions with  
$S_{z}=0$ that one can build using the orbitals with  $\e(k)<0$ and 
half of those with $\e(k)=0$ are a set of  $\left(\begin{array}{c} 2N-2 \\ N-1 \end{array}\right)^{2}$
elements. Each represents one of the  degenerate unperturbed ($U=0$) ground state 
configurations at half filling. First-order perturbation theory requires the diagonalization of 
the $W$ matrix over such a basis.

The diagonal elements of the $W$ matrix are just expectation values over 
determinants $|k_{\a}\ua k_{\b} \da \ldots \ket$.
Such an expectation value is a  sum over all the possible pairs of the 
bielectronic elements of $W$ like 
\begin{eqnarray}
W(\a \b,\a \b)=\sum_{r}U\bra k_{\a}|n_{r}|k_{\a}\ket\bra 
k_{\b}|n_{r}|k_{\b}\ket \nonumber\\
=
\sum_{r}U\frac{1}{N^{2}}e^{i(k_{\a}-k_{\a})r}\frac{1}{N^{2}}e^{i(k_{\b}-k_{\b})r}=
\frac{U}{N^{4}}N^{2}=\frac{U}{N^{2}};
\end{eqnarray}
the result is independent of   $k_{\a}$ and 
$k_{\b}$. Since in any determinant of the set $N^{2}/2$ plane wave states are occupied for each 
spin, there are 
$N^{4}/4$ pairs, and the diagonal elements are all equal to  $UN^{2}/4$. 
Thus, the diagonal elements shift all the eigenvalues by this fixed 
amount.

The off-diagonal elements of the $W$ matrix between determinants that 
differ by three or more spin-orbitals vanish because $\hat{W}$ is a two-body 
operator. The off-diagonal elements  between determinants that 
differ by one spin-orbital are sum of contributions like 
$W(\a \b,\gamma \b)=\sum_{r}U\bra k_{\a}|n_{r}|k_{\g}\ket\bra 
k_{\b}|n_{r}|k_{\b}\ket$ that vanish because of the orthogonality of the 
plane-wave orbitals. One is left with  the off-diagonal elements  between determinants that 
differ by two spin-orbitals, which coincide with the corresponding bielectronic 
elements  $W(\a \beta,\gamma \delta)=\sum_{r}U\bra k_{\a}|n_{r}|k_{\g}\ket\bra 
k_{\b}|n_{r}|k_{\d}\ket$.  This is just  the matrix of $W$ over 
the truncated Hilbert space ${\cal H}$ spanned by the states of the 
holes in the half filled shell, ignoring the filled ones. We stress 
that  there are $N-1$ 
holes of each spin in ${\cal S}_{hf}$, thus ${\cal H}$ is much 
smaller than the full Hilbert space of the Hubbard Hamiltonian; 
however, since the  number of holes grows linearly with $N$, the 
problem is still far from trivial.

\section{Eigenvalues of the number operator}
\label{detdin}

Here we prove Equation (\ref{det}). Expanding the determinant according to its definition
in terms of the totally antisymmetric tensor $\e$, 
\begin{equation}
det|(n_{r})_{ij}-\l\d_{ij}|=\sum_{i_{1}\ldots i_{|{\cal S}|}}\e_{i_{1}\ldots i_{|{\cal S}|}}
(\frac{1}{N^{2}}e^{i(k_{1}-k_{i_{1}})r}-\l\d_{1,i_{1}})\cdot\ldots
\cdot(\frac{1}{N^{2}}e^{i(k_{|{\cal S}|}-k_{i_{|{\cal S}|}})r}-
\l\d_{|{\cal S}|,i_{|{\cal S}|}})
\label{detnx}
\end{equation}
we see that the term of maximum order in $\l$ is $(-\l)^{|{\cal S|}}$; 
it arises from the fundamental permutation 
$(i_{1},i_{2},\ldots)\equiv(1,2,\ldots)$.
The $(|{\cal S}|-1)$-th order term in $\l$ is the sum of $|{\cal S}|$ 
identical contributions also arising from the fundamental permutation. 
Therefore, it is
\begin{equation}
\frac{|{\cal S}|}{N^{2}}(-\l)^{|{\cal S}|-1}.
\end{equation}
It is not difficult to see that all the other orders in $\l$ yield nothing. 
At order zero one finds
\begin{equation}
\frac{1}{N^{2|{\cal S}|}}\sum_{i_{1}\ldots i_{|{\cal S}|}}\e_{i_{1}\ldots i_{|{\cal S}|}}
e^{i(k_{1}+\ldots+k_{|{\cal S}|}-
k_{i_{1}}-\ldots-k_{i_{|{\cal S}|}})r}.
\label{zeroord}
\end{equation}
Since the exponential is totally symmetric in the permutation of 
$i_{1} \ldots i_{|{\cal S}|}$ while  $\e$ is totally antisymmetric the 
sum vanishes. 

Now we analize the first order term in $\l$. One of its contributions 
is obtained by picking $-\l$ in the first factor of eq.(\ref{detnx}), 
i.e. by setting $i_{1}=1$.  This contribution can be written as
\begin{equation}
\frac{1}{N^{2(|{\cal S}|-1)}}\sum_{i_{2}\ldots i_{|{\cal S}|}}\e_{1i_{2}\ldots i_{|{\cal S}|}}
e^{i(k_{2}+\ldots+k_{|{\cal S}|}
-k_{i_{2}}-\ldots-k_{i_{|{\cal S}|}})r}.
\label{firstord}
\end{equation}
Again, the exponential is symmetric $i_{2} \ldots i_{|{\cal S}|}$ 
and the sum in eq.(\ref{firstord}) vanishes. 
Clearly, this argument applies to all the contributions to first order and to 
all orders less than  $|{\cal S}|-1$.

So, we are left with the only two nonvanishing terms,
\begin{equation}
det|n_{r}-\l|=(-\l)^{|{\cal S}|-1}(\frac{|{\cal S}|}{N^{2}}-\l)
\end{equation}
that is,  eq.(\ref{det}).

\section{ Spin Projection Operators $P_{S}$}
\label{spin}

Let us consider a determinantal state with $2n$  spin-orbitals, half 
of spin up and half of spin down, like \cite{caso}
\begin{equation}
    |\Psi\ket_{\s} \equiv |\q_{1}\ldots\q_{n}\ket_{\s}
    |\q_{n+1}\ldots\q_{2n}\ket_{-\s}
\label{genericstate}
\end{equation}
with  $\bra\q_{i}|\q_{j}\ket=\d_{ij}$.  Due to the anticommuting property of the fermionic 
operators, $|\q\ket$ is separately antisymmetric in the 
indices $1,\ldots,n$ and $n+1,\ldots,2n$. $|\Psi\ket_{\s}$ is no 
eigenstate of $\hat{S}^{2}$ and $\forall S=0,\ldots 
n$, $P_{S}|\Psi\ket_{\s}\neq 0$. To build $P_{S}$, we take advantage of the 
fact that different $S$ correspond to different irreps of the 
Permutation Group of the spins. To this end, let us 
draw the following table, or Young Tableau \cite{greiner}
\begin{equation}
\begin{array}{l}
\begin{tabular}{|c|c|c|c|c|c|c|}
\hline 
$i_{1}$  & $i_{2}$ & $\ldots$ & $i_{n-S}$ & $i_{n-S+1}$  & $\ldots$ & 
$i_{n+S}$ \\
\hline 
\end{tabular}
\\
\begin{tabular}{|c|c|c|c|}
$j_{1}$ & $j_{2}$ & $\ldots$ & $j_{n-S}$  \\
\hline  
\end{tabular}
\end{array}
\label{table}
\end{equation}
where the values of the indices $i_{k}$ and $j_{k}$ must be a permutation of 
$1,\ldots,2n$ with the constraints 1)  $i_{k}<j_{k}\;\forall 
k=1,\ldots,n-S$ and 2) $i_{k}<i_{k+1}$, $j_{k}<j_{k+1}$. Now we define 
$C_{i,j}$ as the operator that exchanges the spins of the $|\q_{i}\ket$ 
and  $|\q_{j}\ket$ states; we associate to 
the above table the operator
\begin{equation}
    \prod_{a=1}^{n+S-1}(1+\sum_{\a=1}^{n+S-a}C_{i_{\a},i_{n+S-a+1}})
\prod_{b=1}^{n-S-1}(1+\sum_{\b=1}^{n-S-b}C_{j_{\b},j_{n-S-b+1}})
\prod_{l=1}^{n-S}(1-C_{i_{l},j_{l}}).
\end{equation}
This operator antisymmetrizes the indices on the same column and then 
symmetrizes those on the same row. Then $P_{S}$ is proportional, up to a 
normalization factor, to the sum of the operators associated to all  
tables (\ref{table}) that comply with the constraints 1) and 2). 

Example: $n=2 \Longrightarrow|\Psi\ket_{\s}=
|\q_{1}\ua\q_{2}\ua\q_{3}\da\q_{4}\da\ket\equiv |\q_{1}\q_{2}\ket_{\ua}
    |\q_{3}\q_{4}\ket_{\da}$. In this case the three 
projection operators are
\begin{equation}
\begin{array}{l}
P_{S=2}\propto
\begin{tabular}{|c|c|c|c|}
\hline
   1 & 2 & 3 & 4 \\
   \hline
\end{tabular}
\\
\\
P_{S=1}\propto
\begin{array}{l}
    \begin{tabular}{|c|c|c|}
\hline
   1 & 2 & 3  \\
   \hline
   \end{tabular}
\\
\begin{tabular}{|c|}
4 \\
\hline  
\end{tabular}
\end{array}+
\begin{array}{l}
    \begin{tabular}{|c|c|c|}
\hline
   1 & 2 & 4  \\
   \hline
   \end{tabular}
\\
\begin{tabular}{|c|}
3 \\
\hline  
\end{tabular}
\end{array}+
\begin{array}{l}
    \begin{tabular}{|c|c|c|}
\hline
   1 & 3 & 4  \\
   \hline
   \end{tabular}
\\
\begin{tabular}{|c|}
2 \\
\hline  
\end{tabular}
\end{array}
\\
\\
P_{S=0}\propto
\begin{array}{l}
    \begin{tabular}{|c|c|}
\hline
   1 & 2   \\
   \hline
   \end{tabular}
\\
\begin{tabular}{|c|c|}
3 & 4 \\
\hline  
\end{tabular}
\end{array}+
\begin{array}{l}
    \begin{tabular}{|c|c|}
\hline
   1 & 3   \\
   \hline
   \end{tabular}
\\
\begin{tabular}{|c|c|}
2 & 4 \\
\hline  
\end{tabular}
\end{array}.
\end{array}
\end{equation}
If a table has two indices corresponding to states with equal spin 
in the same column, the action of the associated 
operator on $|\Psi\ket_{\s}$ will yield zero. This 
means that in the above example we can omit the third table of 
$P_{S=1}$ and the second one of $P_{S=0}$. We have special intererst 
in $P_{S=0}$, which projects on the totally antisymmetric irrep. For all  
$n$, it  always consists of only one table 
\begin{equation}
P_{S=0}=
\begin{array}{l}
    \begin{tabular}{|c|c|c|c|}
\hline
   1 & 2  & \ldots & $n$ \\
   \hline
  $n+1$ & $n+2$ & \ldots & $2n$ \\
   \hline
   \end{tabular}
\end{array}.
\end{equation}
Using the explicit form of $P_{S=0}$ one finds that 
 the singlet projection contains $|\Psi\ket_{\s}$ and $|\Psi\ket_{-\s}$ 
in a  combination, which is symmetric  for even $n$ and  antisymmetric  if $n$ 
is  odd.


\begin{references}
\bibitem{appendix} $C_{4v}$ is the symmetry group of a square. 
It is a finite group of 
order 8 and it contains 4 one dimensional irreps, 
$A_{1},\;A_{2},\;B_{1},\;B_{2}$, and 1 two-dimensional one called $E$. 
The table of characters is
\begin{center}
    \vspace*{0.5 cm}
\begin{tabular}{|c|c|c|c|c|c|}
\hline 
$C_{4v}$ & $\mathbf{1}$ & $C_{2}$ & $C^{(+)}_{4},\,C^{(-)}_{4}$ & 
$\s_{x},\,\s_{y}$ & 
$\s'_{+},\,\s'_{-}$ \\
\hline 
$A_{1}$ & 1 & 1 & 1 & 1 & 1 \\
\hline 
$A_{2}$ & 1 & 1 & 1 & -1 & -1 \\
\hline 
$B_{1}$ & 1 & 1 & -1 & 1 & -1 \\
\hline 
$B_{2}$ & 1 & 1 & -1 & -1 & 1 \\
\hline 
$E$ & 2 & -2 & 0 & 0 & 0 \\
\hline 
\end{tabular}
\vspace*{0.5 cm}
\end{center}
\bibitem{he} Equivalently, we shall find the exact $2N-2$-body ground state of the  
effective hamiltonian 
\begin{equation}
H_{eff}=\frac{U}{N^{4}}\sum_{k_{1},k_{2},k_{3},k_{4}\in {\cal S}_{hf}}
\d(k_{1}+k_{2}-k_{3}-k_{4})
c^{\dag}_{k_{1}\ua}c^{\dag}_{k_{2}\da}c_{k_{3}\da}c_{k_{4}\ua}
\label{heff}
\end{equation}
where $\d(G)=1$ if $G$ is a reciprocal lattice vector and zero 
otherwise.
\bibitem{caso} We recall that the $W=0$ state $|\F_{AF}\ket_{\s}$ is of this form, with  $n=N-1$.
\end{references}
\end{document}